# The Design of Efficiently-Encodable Rate-Compatible LDPC Codes

Jaehong Kim, Aditya Ramamoorthy, *Member*, *IEEE*, and Steven W. McLaughlin, *Fellow, IEEE*

**Abstract**

We present a new class of irregular low-density parity-check (LDPC) codes for moderate block lengths (up to a few thousand bits) that are well-suited for rate-compatible puncturing. The proposed codes show good performance under puncturing over a wide range of rates and are suitable for usage in incremental redundancy hybrid-automatic repeat request (ARQ) systems. In addition, these codes are linear-time encodable with simple shift-register circuits. For a block length of 1200 bits the codes outperform optimized irregular LDPC codes and extended irregular repeat-accumulate (eIRA) codes for all puncturing rates 0.6~0.9 (base code performance is almost the same) and are particularly good at high puncturing rates where good puncturing performance has been previously difficult to achieve.

## I. INTRODUCTION

Low-density parity-check (LDPC) codes are considered good candidates for next-generation forward error control in high throughput wireless and recording applications. Their excellent performance and parallelizable decoder make them appropriate for technologies such as DVB-S2, IEEE 802.16e, and IEEE 802.11n. While semiconductor technology has progressed to an extent where the implementation of LDPC codes has become possible, many issues still remain. First and foremost, there is a need to reduce complexity without sacrificing performance. Second, for applications such as wireless LAN, the system throughput depends upon the channel conditions

Jaehong Kim is with the department of Electrical & Computer Engineering, Georgia Institute of Technology, Atlanta, GA USA (e-mail: onil@ece.gatech.edu).
Aditya Ramamoorthy is the department of Electrical & Computer Engineering, Iowa State University, Ames, IA, USA (e-mail: adityar@iastate.edu)
Steven W. McLaughlin is with the department of Electrical & Computer Engineering, Georgia Institute of Technology, Atlanta, GA USA (e-mail: swm@ece.gatech.edu).

and hence the code needs to have the ability to operate at different rates. Third, while the LDPC decoder can operate in linear time, it may be hard to perform low-complexity encoding of these codes. While the encoding time of irregular LDPC codes can be reduced substantially using the techniques presented in [1] at long block lengths, their techniques may be hard to apply at short block lengths. The other option is to resort to quasi-cyclic (QC) LDPC constructions that can be encoded by shift registers [2]. However such constructions are typically algebraic in nature and usually result in codes with regular degree distributions.

An important problem is the design of LDPC codes that can be easily encoded and have good puncturing performance across a wide range of rates. In this work, we introduce a new class of LDPC codes called Efficiently-Encodable Rate-Compatible ($E^2RC$) codes that have a linear-time encoder and have good performance under puncturing for a wide variety of rates. Section II overviews prior work in irregular LDPC codes and rate-compatible puncturing. In section III, we present the $E^2RC$ construction algorithm. The shift-register based encoder structure for the $E^2RC$ codes is explained in section IV. Section V compares the puncturing performance of the $E^2RC$ codes with that of other irregular LDPC codes and section VI outlines the conclusions.

## II. Background and Related Work

LDPC codes can be defined by a sparse binary parity-check matrix of size $M \times N$, where $M$ and $N$ are the number of parity symbols and codeword symbols respectively. The parity-check matrix can equivalently be considered as a bipartite graph (called the Tanner graph of the code [3]), where columns and rows in the parity-check matrix correspond to variable nodes and check nodes on the graph, respectively. The distribution of variable (check) nodes in the graph can be represented as a polynomial $\lambda(x) = \sum \lambda_i x^{i-1}$ ($\rho(x) = \sum \rho_i x^{i-1}$), where $\lambda_i$ ($\rho_i$) is the fraction of edges incident to variable (check) nodes of degree $i$.

In this paper we work with systematic LDPC codes. Thus, assuming that the parity-check matrix is full-rank, we have $K$ columns corresponding to information bits and $M$ columns corresponding to parity bits, where $K + M = N$. In the sequel we shall refer to the submatrix of the parity-check matrix corresponding to the $K$ information bits, the systematic part and the submatrix corresponding to the parity bits, the nonsystematic part. We shall denote the parity-check matrix by $H$, the systematic part by $H_1$ and the nonsystematic part by $H_2$. Thus, $H = [H_1 \,|\, H_2]$.

*A. Extended Irregular Repeat Accumulate Codes*

A promising class of LDPC codes called Irregular Repeat Accumulate (IRA) codes was introduced by Jin et al. in [4]. These codes have several desirable properties. First, IRA codes can be encoded in linear time like Turbo codes. Second, their performance is superior to turbo codes of comparable complexity and as good as best known irregular LDPC codes [4]. The columns corresponding to the degree-two nodes in the parity check matrix of IRA codes has bi-diagonal structure shown below (1).

$$H_2 = \begin{bmatrix} 1 & & & & & \\ 1 & 1 & & & & \\ & 1 & 1 & & & \\ & & & \ddots & & \\ & & & & 1 & 1 & \\ & & & & & 1 & 1 \end{bmatrix}, \qquad (1)$$

The class of extended IRA (eIRA) codes was introduced by Yang *et al.* in [5]. The eIRA codes achieve good performance by assigning degree-2 nodes to nonsystematic bits and ensuring that the degree-2 nodes do not form a cycle amongst themselves. Furthermore, they avoid cycles of length four and make the systematic bits correspond to variable nodes of degree higher than two. They ensure efficient encoding by forming the parity in the bi-diagonal structure like IRA codes as shown in (1). For more details we refer the reader to [5].

It is interesting to see whether there exist other ways of placing the degree-2 nodes without any cycles involving only degree-2 nodes. We present an example of such a placement below in the case when $M = 8$.

$$H_2 = \begin{bmatrix} 1 & 0 & 0 & 0 & 0 & 0 & 0 & 0 \\ 0 & 1 & 0 & 0 & 0 & 0 & 0 & 0 \\ 0 & 0 & 1 & 0 & 0 & 0 & 0 & 0 \\ 0 & 0 & 0 & 1 & 0 & 0 & 0 & 0 \\ 1 & 0 & 0 & 0 & 1 & 0 & 0 & 0 \\ 0 & 1 & 0 & 0 & 0 & 1 & 0 & 0 \\ 0 & 0 & 1 & 0 & 1 & 0 & 1 & 0 \\ 0 & 0 & 0 & 1 & 0 & 1 & 1 & 1 \end{bmatrix}. \quad (2)$$

Observe that the column degree of each node is 2 except the last column and that there does not exist any cycle in this matrix. We shall see later that this construction can be generalized and the resulting matrices can be used to construct LDPC codes that can be efficiently encoded and have good puncturing performance over a wide range of rates.

### *B. Rate-Compatible Puncturing*

In wireless channels where the channel conditions vary with time, using systematic codes and puncturing the parity bits is an efficient strategy for rate-adaptability, since the system requires only one encoder-decoder pair. Rate-compatible punctured codes (RCPC) were introduced by Hagenauer [6] as an efficient channel coding scheme for Incremental Redundancy (IR) Hybrid-Automatic Repeat reQuest (ARQ) schemes. In an RCPC family, the parity bits of a higher-rate code are a subset of the parity bits of the lower-rate code. If the receiver cannot decode the code based on the current received bits, it requests the transmitter for additional parity bits until it decodes correctly. Thus, the subset property that the parity bits of codes of different rate satisfy is useful.

Rate-compatible puncturing of LDPC codes was considered by Ha *et al.* [7]. They derived the density evolution equations for the design of good puncturing degree distributions under the Gaussian Approximation. Ha *et al.* also proposed an efficient puncturing algorithm for a given mother code in [8-9]. For finite length (up to several thousand symbols) LDPC codes, Yazdani *et al*. construct rate-compatible LDPC codes using puncturing and extending [10].

The algorithm of [8-9] takes as input a particular mother code Tanner graph and a set of target rates. It then performs a search to identify the set of codeword symbols that should be punctured to achieve those target rates. Since our code construction technique is inspired by it, we present a brief description of the algorithm below.

Suppose that the Tanner graph of the mother code is denoted by $G = (V \bigcup C, E)$, where $V$ denotes the set of variable nodes, $C$ denotes the set of check nodes and $E$ denotes the set of edges. Let $S \subseteq V$ be a subset of the variable nodes. Then the set of check node neighbors of $S$ shall be denoted by $\mathcal{N}(S)$. Similar notation shall be used to denote the set of variable node neighbors of a subset of the check nodes. The set of unpunctured nodes is denoted by $V_0$ and the set of punctured variable nodes is denoted by $V \setminus V_0$ (using standard set-theory notation).

*Definition 1* [**1-step recoverable node]:** A punctured variable node $p \in V \setminus V_0$ is called a *1*-step recoverable (*1*-SR) node if there exists $c \in \mathcal{N}(\{p\})$ such that $\mathcal{N}(\{c\}) \setminus \{p\} \subseteq V_0$.

*1*-step recoverable nodes are so named because in the absence of any channel errors these nodes can be decoded in one step of iterative decoding. This definition can be generalized to *k*-step recoverable (*k*-SR) nodes (see Fig. 1). Let $V_1$ be the set of *1*-SR nodes among the punctured variable nodes. Similarly, let $V_k$ be the set of *k*-step recoverable nodes, which are defined as follows:

*Definition 2*: A punctured variable node $p \in V \setminus V_0$ is called *k*-step recoverable (*k*-SR) node if there exists $c \in \mathcal{N}(\{p\})$ such that $\mathcal{N}\left(\{c\}\right) \setminus \{p\} \subseteq \bigcup_{i=0}^{k-1} V_i$ and that there exists $q \in \mathcal{N}\left(\{c\}\right) \setminus \{p\}$, where $q \in V_{k-1}$.

From the above two definitions, note that $V = \bigcup_{i=0}^{\infty} V_i$ ($V_\infty$ represents the set of nodes that cannot be recovered by erasure decoding). Under these conditions, note that the *k*-SR node will be recovered after exactly *k* iterations of iterative decoding assuming that the channel does not cause any errors. So a large number of low-SR nodes are intuitively likely to reduce the overall number of iterations, which results in good puncturing performance. The general idea of the puncturing algorithm in [8-9] is to find a good ensemble of sets $V_k$'s maximizing $|V_k|$ for small $k > 0$. The puncturing algorithm in [8-9] works well for any given mother code. However, the maximum puncturing rate is often limited when this algorithm is applied, so that high puncturing rates are difficult to achieve. This is because it is difficult to find enough number of low-SR nodes from a randomly constructed matrix. In addition, [8-9] do not address the problem of mother code design for puncturing, i.e., they do not present a technique for the design of a mother code in which the parity check matrix has a large number of variable nodes that are $k$-step recoverable with low values of $k$. This is the focus of this paper.

## III. A New Class of Irregular LDPC Codes

In this work we are interested in designing rate-compatible punctured codes that exhibit good performance across a wide range of coding rates. To ensure good performance over the different coding rates we attempt to design the mother code matrix to have a large number of $k$-SR nodes with low values of $k$. From a practical perspective the requirement of low-complexity encoding is also important. Like punctured RA, IRA and eIRA codes, these codes are designed to recover all the punctured bits when the channel is error-free even when they achieve the maximum puncturing rate by running sufficient iterations of iterative decoding. Thus, encoding of these codes is also relatively simple.

### *A. Code Construction Algorithm*

Before describing our design algorithm, we define a $k$-SR matrix. Let $v_i$ denote the $i$-th column of the parity-check matrix $H$, where $0 \le i < N$. We shall use it interchangeably to denote the variable node corresponding to the $i$-th column in the Tanner graph of $H$.

*Definition 3*: The matrix $P = (v_s)_{s \in S}$ is called a $k$-SR matrix, if $v_s \in V_k$ for all $s \in S$, where $S \subseteq \{0,1,\cdots,N-1\}$.

In the proposed E$^2$RC codes, we construct the parity-check matrix by placing several $k$-SR matrices as shown in Fig. 2. We assign all the degree-2 nodes to the nonsystematic part. Nodes having degree higher than two are elements of the *0*-SR matrix which consists of message nodes and parity nodes that shall not be punctured. Consider the submatrix of *0*-SR matrix formed by the high degree nodes in the nonsystematic part. We denote such submatrix of *0*-SR matrix as *L*, and the number of columns in *L* as *l* as depicted in Fig. 2(a).

*Definition 4*: The depth *d* is the number of different types of *k*-SR matrices that have degree-2 columns in a parity-check matrix.

*Definition 5*: The function $\gamma(k)$ is the number of columns in the *k*-SR matrix in a parity-check matrix, i.e., $\gamma(k) = |V_k|$, where $k > 0$.

From Definition 5, note that the size of the *k*-SR matrix is $M \times \gamma(k)$. Let $N_v(i)$ represent the number of variable nodes of degree $i$. Fig. 2(a) shows the case when $N_v(2) < M - 1$, and we shall elaborate on the design of such codes in subsection III.B. Other than that, we assume that $N_v(2) = M - 1$ throughout the paper. When $N_v(2) > M - 1$ we cannot guarantee the cycle-free property among the degree-2 nodes, which is an important design rule that will be explained later. When $N_v(2) = M - 1$, there will be no *0*-SR nodes in the nonsystematic part, i.e. $l = 0$. In this case, we insert a degree-1 node in the last column of nonsystematic part, and assign all the variable nodes of the nonsystematic part to degree-2 nodes except the last degree-1 node as shown in Fig. 2(b).

*Example 1*: For $M = 8$ and $N_v(2) = 7$, we can construct the nonsystematic part $H_2$ as in (2). In (2), the first four columns form the *1*-SR matrix, the next two columns form the *2*-SR matrix, and the next one column forms the *3*-SR matrix. Thus, depth $d = 3$, $\gamma(1) = 4$, $\gamma(2) = 2$, and $\gamma(3) = 1$. We can also regard the last degree-1 column as *4*-SR matrix. However, our convention in this paper is to only consider degree-2 columns to calculate the depth *d*. From now on, we refer to the last degree-1 column in $H_2$ as (*d+1*)-SR matrix since the connections with other *k*-SR matrices makes it (*d+1*)-SR node. ■

Let $S_k = \sum_{j=1}^{k} \gamma(j)$. Thus, $S_k$ represents the sum of the number of columns in the submatrix formed by the placing the *1*-SR, *2*-SR, … and *k*-SR matrices next to each other. We set $S_0$ to 0. We shall represent the position of the ones in a column belonging to a *k*-SR matrix by the powers of a polynomial in $D$. According to our construction, the $j$-th column of $k$-SR matrix can be represented by the following polynomial

$$h_{k,j} = D^{j+S_{k-1}}\left(1+D^{\gamma(k)}\right), \quad \text{where} \quad 1 \le k \le d, 0 \le j \le \gamma(k)-1 \ \text{and}$$

$$h_{d+1} = D^{M-1}.$$

In the sequence, $D^i$ represents the position of nonzero element in a column, i.e., $i$-th element of the column is nonzero, where $0 \le i \le M-1$. For Example 1, we note that the depth can be obtained by setting $d = \log_2 M = \log_2 8 = 3$ and $\gamma(k) = M/2^k$ for $1 \le k \le d$, $\gamma(d+1) = 1$. In general, $M$ need not be a power of two. We present the algorithm for constructing $H_2$ for general $M$ below.

## $E^2$RC Code Construction Algorithm

STEP 1 [**Finding Optimal Degree Distribution**] Find an optimal degree distribution for the desired code rate with constraint that $N_v(2) < M$.

STEP 2 [**Parameter Setting**] For a given design parameter, $M$ (number of parity symbols), obtain the depth $d$ and $\gamma(k)$. The computation of $d$ and $\gamma(k)$ is explained below. The size of the $k$-SR matrix is set to be $M \times \gamma(k)$.

STEP 3 [**Generating *k*-SR matrix**] The $j$-th column of the $k$-SR matrix has the following sequence:

$$h_{k,j} = \begin{cases} D^{j+S_{k-1}}\left(1+D^{\gamma(k)}\right), & for \quad 1 \le k \le d \\ D^{M-1}, & for \quad k = d+1 \end{cases}, \quad where \quad 0 \le j \le \gamma(k)-1.$$

STEP 4 [**Constructing matrix *T***] Construct the matrix $T$ as follows:

$$T = \left[1\text{-SR matrix} \mid 2\text{-SR matrix} \mid \cdots \mid d\text{-SR matrix}\right].$$

STEP 5 [**Forming matrix *H₂***] Add a degree-1 node to $T$ and form $H_2 = \left[T \mid (d+1)\text{-SR matrix}\right]$.

STEP 6 [**Edge Construction**] Construct the matrix $H_1$ by matching the degree distribution obtained in STEP 1 as closely as possible.

STEP 7 [**Constructing matrix *H***] Assign $H_1$ as the systematic part and $H_2$ as the nonsystematic part:

$$H = \left[H_1 \mid H_2\right].$$

In STEP 1, we first find an optimal degree distribution for the desired mother code rate, say $R_L$, using the density evolution [11]. When we determine the degree distribution, the number of degree-2 nodes, $N_v(2)$, is an important factor. The E[2]RC codes are designed so that all the degree-2 nodes in the nonsystematic part can be punctured. This will give us the highest achievable puncturing rate, say $R_H$. Then, $R_H = K/(N - N_v(2))$. Thus, the E[2]RC codes can provide an ensemble of rate-compatible codes of rate $R_L \sim R_H$. When $N_v(2) = M - 1$ all the parity bits have degree two and can be punctured so that $R_H = 1.0$. In STEP 2, we set the design parameters. We try to obtain a large number of low-SR nodes while constraining the increase in the row degree. In fact, we design the function $\gamma(k)$ such that it assigns approximately half of the parities as *1*-SR nodes, and approximately the half of the remaining parities as *2*-SR nodes, and so on. The depth $d$ is given as $d = \lceil \log_2 M \rceil$, and $\gamma(k)$ as

$$\gamma(k) = \left\lfloor M - \frac{1}{2}\sum_{i=0}^{k-1} \gamma(i) \right\rfloor \text{ for } 1 \le k \le d,\ \gamma(d+1) = 1, \text{ and } \gamma(0) \triangleq M \tag{3}$$

where $\lceil \cdot \rceil$ and $\lfloor \cdot \rfloor$ are the ceiling function and the floor function, respectively. We observe that the function $\gamma(k)$ is such that the following facts hold true.

*Fact 1*: The function $\gamma(k)$ is such that $S_d = \sum_{i=1}^{d} \gamma(i) = M - 1$, where $d = \lceil \log_2 M \rceil$. Furthermore $\gamma(k) \ge 1$ for $1 \le k \le d$.

*Proof*: See Appendix. ■

From the generation sequence in STEP 3, we can notice that the $k$-SR matrix is composed of only degree-2 variable nodes except for the last ($d$+*1*)-SR matrix. Note that every column in $k$-SR matrix has degree two. In particular, when $N_v(2) = M - 1$, all the columns of the nonsystematic part have degree two except the last column which has degree one.

After generating the $k$-SR matrices, we put them together to form the matrix $T$ in STEP 4. Then in STEP 5, we construct the nonsystematic part $H_2 = [T \mid (d+1)\text{-SR matrix}]$ by adding a degree-1 column at the end of $H_2$. Example 2 shows an example of the construction of a $H_2$ matrix using the proposed algorithm.

*Example 2*: For $M = 7$ and $N_v(2) = 6$, the depth $d = 3$, and $\gamma(1) = 3$, $\gamma(2) = 2$, $\gamma(3) = 1$.

$$H_2 = \begin{bmatrix} 1 & 0 & 0 & 0 & 0 & 0 & 0 \\ 0 & 1 & 0 & 0 & 0 & 0 & 0 \\ 0 & 0 & 1 & 0 & 0 & 0 & 0 \\ 1 & 0 & 0 & 1 & 0 & 0 & 0 \\ 0 & 1 & 0 & 0 & 1 & 0 & 0 \\ 0 & 0 & 1 & 1 & 0 & 1 & 0 \\ 0 & 0 & 0 & 0 & 1 & 1 & 1 \end{bmatrix}$$

■

In STEP 6 the matrix $H_1$ is constructed by trying to match the degree distribution obtained from STEP 1. Note that the degree distribution of the nonsystematic part is already fixed by the construction algorithm. This may cause some check nodes to have degrees higher than those specified by the optimal degree distribution. In this case we try to match the optimal degree distribution as closely as possible. Since we have some high degree check nodes, we compensate it to match the average right degree by enlarging the number of lower degree check nodes or placing some lower degree check nodes. Finally, $H_1$ and $H_2$ are combined to make the whole parity-check matrix in STEP 7.

We now present some properties of the codes that are constructed using the previous algorithm. In the subsequent statements and discussion, unless otherwise specified, $H_2$ shall represent the nonsystematic part of a parity-check matrix and shall be assumed to have been generated by the $E^2$RC construction algorithm.

*Lemma 1:* In the matrix $H_2$, any column in a $k$-SR matrix is connected to at least one row of degree-$k$. Furthermore, this row has exactly one connection to a column from each $l$-SR matrix, where $1 \le l < k \le d$.

*Proof:* See Appendix.

From Lemma 1, it is possible to find the exact number of rows with degree-$k$ except the last row. We define $\zeta$ as the row degree of the last row.

*Observation 1:* The row degree $\zeta$ of the last row in the matrix $H_2$ can be obtained as $\zeta = \sum_{i=1}^{d} \left[ \gamma(i) + S_i - S_d \right] + 1$.

*Proof*: Consider the connections of the last row with each $k$-SR matrix. It is easy to see that if $M = 2\cdot\gamma(1)$, there is a connection between the *1*-SR matrix and the last row, otherwise, there is no connection. Similarly, if $M = \gamma(1)+2\cdot\gamma(2)$, there is a connection between the *2*-SR matrix and the last row, and so on. Thus, we can get $\zeta$ as

$$\zeta = \underbrace{\big(1-(M-2\gamma(1))\big)}_{\textit{1}\text{-SR matrix}} + \underbrace{\big(1-(M-\gamma(1)-2\gamma(2))\big)}_{\text{2-SR matrix}} + \cdots + \underbrace{\big(1-(M-\gamma(1)-\gamma(2)-\cdots-2\gamma(d))\big)}_{d\text{-SR matrix}} + \underbrace{1}_{(d+\textit{1})\text{-SR matrix}}$$

$$= \sum_{i=1}^{d}\left[\gamma(i) + S_i - (M-1)\right] + 1$$

$$= \sum_{i=1}^{d}\left[\gamma(i) + S_i - S_d\right] + 1,$$

since we have $S_d = M-1$ from Fact 1. ■

From Observation 1, we can obtain $\zeta = \sum_{i=1}^{3}\left[\gamma(i) + S_i - 6\right] + 1 = 3$ for Example 2. Since we know $\zeta$, we are ready to get the right degree distributions for $H_2$.

*Observation 2*: The number of degree-$k$ rows in the matrix $H_2$ is $\gamma(k) + \delta(k-\zeta)$ for $1 \le k \le d$, where $\delta(i) = \begin{cases} 1 & \textit{if } i = 0, \\ 0 & \textit{otherwise} \end{cases}$.

*Corollary 1:* The right degree distribution (node perspective) of the matrix $H_2$ is as follows:

$$\rho(x) = \sum_{i=1}^{d+1} \hat{\rho}_i x^{i-1}, \text{ where } \hat{\rho}_i = \frac{\gamma(i) + \delta(i-\zeta)}{M} \text{ for } 1 \le k \le d \text{ and } \hat{\rho}_{d+1} = \frac{\delta(i-\zeta)}{M}.$$

*Proof:* Consider the $k$-SR matrix when $1 \le k \le d$. From Lemma 1, if we pick a column in the $k$-SR matrix, the first element of the column is included in a row of degree $k$, and the second element has row degree greater than $k$. The number of columns in the $k$-SR matrix is $\gamma(k)$ and each column is connected to one degree-$k$ row. Thus, the number of rows having degree $k$ is at least $\gamma(k)$ except the last row. For a ($d$+*1*)-SR matrix, there is only one degree-$\zeta$ row. From Fact 1, summing the number of rows having degree-$k$ results in $\gamma(1) + \gamma(2) + \cdots + \gamma(d) + 1 = M$. Therefore, the number of rows of degree $k$ except the last row is exactly $\gamma(k)$. The result follows. ■

Once the optimal degree distributions for the whole code for a desired code rate have been found, we can get the degree distributions for the $H_1$ matrix while fixing the degree distributions obtained from the construction algorithm for $H_2$. In general, matching the optimal degree distribution for the whole code may not be possible because of the construction algorithm. For the systematic part, namely the $H_1$ matrix, we choose variable nodes of higher degree greater than two. Besides finding the optimal degree distributions, there are three additional design rules for finite-length LDPC codes proposed in [11]:

(a) Assign degree-2 variable nodes to nonsystematic bits;

(b) Avoid short cycles involving only degree-2 variable nodes.

(c) Avoid cycles of length four.

The proposed $E^2$RC codes meet the design rule (a) as stated above. For design rule (b), we show that there are no cycles involving only degree-2 variable nodes.

*Lemma 2:* Suppose that there exists a length-2$s$ cycle in a matrix which consists of only weight two columns. Consider the submatrix formed by the subset of columns that participates in the cycle. Then, all the participating rows in the cycle must have degree two in that submatrix.

*Proof:* To have a length-2$s$ cycle, the number of columns participating in the cycle needs to be $s$ and the number of rows participating in the cycle needs to be $s$. Let us denote the submatrix formed by the columns participating in the cycle by $U$. Then, the number of edges in $U$ is 2$s$ since each of the columns has degree two. Each row participating in the cycle must have a degree greater than or equal to two in $U$ since each row has to link at least two different columns in $U$. Suppose there is a row having degree strictly greater than two in $U$. Then there should be a row having degree less than two in $U$, since the average row weight in $U$ is two (the number of edges / the number of rows = 2$s$ / $s$ = 2). This is a contradiction because a row that has degree less than two in $U$ cannot participate in a cycle with the columns in $U$. Thus, every participating row must have degree two in $U$. ■

Using Lemma 2, we prove that the proposed matrix $H_2$ is cycle free.

*Lemma 3:* The matrix $H_2$ constructed by the $E^2$RC construction algorithm is cycle free.

*Proof:* Suppose that there exist $s$ columns $v_1, v_2, \cdots, v_s$ in $H_2$ that form a cycle of length $2s$. We form the $M \times s$ submatrix formed by the columns. Let us denote this submatrix by $H_s$. Suppose that column $v_i$ belongs to the $k_i$-SR matrix in $H_2$. Let $k_{min} = \min_{\{i\}} k_i$. Applying Lemma 1, we have that $v_{k_{\min}}$ has exactly one connection to each $l$-SR matrix, where $1 \le l < k_{min}$, and no connection to $m$-SR matrices where $m > k_{min}$, i.e., there is a check node connected to $v_{k_{\min}}$ that is singly-connected in the submatrix $H_s$. Applying Lemma 2, we realize that a cycle cannot exist amongst the $s$ columns. ■

The matrix $H_2$ has a high fraction of degree-2 nodes. In fact, if $N_v(2) = M-1$ then $(M-1)/M$ fraction of the nodes in $H_2$ are of degree-2, and there is only one degree-1 node. The construction algorithm also induces a spread in the check node distribution. This may cause the constructed codes to have higher error floors. To reduce these effects, we can use methods such as those presented in [12-15] when we construct the $H_1$ matrix. By doing so, the E$^2$RC codes can meet the design rule (c).

### B. Low-Rate E$^2$RC Code Design

Considering E$^2$RC mother code design for low rate ($R < 0.5$) is a natural step. In this case, we should consider a design that allows some portion of the nodes in the nonsystematic part to have degree greater than two since it is hard to obtain a good degree distribution that has all the parity bits of degree two. This is the reason why we consider the case when $N_v(2) < M-1$. We will briefly explain the differences in the construction algorithm for this case compared to the case considered earlier. Recall that we puncture only the degree two nodes. The matrix $L$ that has $l$ columns shown in Fig. 2(a) consists of those parity bits that have degree higher than two and shall not be punctured. Since $N_v(2) = M - l < M-1$, we set the depth as the maximum $d$ such that $S_{d-1} < N_v(2)$, and obtain $\gamma(k)$ as before for $1 \le k < d$. The previous settings for $\gamma(k)$'s are designed to match $S_d = N_v(2) = M-1$. In this case, however, we set $\gamma(d) = N_v(2) - S_{d-1}$ so that they can satisfy $S_d = N_v(2)$. To generate the sequence of $d$-SR matrix, we set

$$\delta = \left\lfloor M - \frac{1}{2}\sum_{i=0}^{d-1} \gamma(i) \right\rfloor.$$

Then, the $j$-th column of $k$-SR matrix of STEP 3 has the following sequence:

$$h_{k,j} = \begin{cases} D^{j+S_{k-1}}\left(1+D^{\gamma(k)}\right), & for \quad 1 \le k < d \\ D^{j+S_{k-1}}\left(1+D^{\delta}\right), & for \quad k = d \end{cases}, \quad where \quad 0 \le j \le \gamma(k)-1 .$$

We formulate $T$ in the same way as before and set $H_2 = [L \mid T]$, where variable nodes in the matrix $L$ have degree higher than two. Note that we do not put the degree-1 node in $H_2$. Then, we need to construct edges for the matrix $L$ and $H_1$ by trying to match the target degree distribution and by avoiding cycles of length four. Note that the submatrix formed by the columns of $N_v(2)$ is cycle free (the proof is very similar to the previous proof).

For the proposed codes, rate-compatibility can be easily obtained by puncturing the degree-two nodes from left to right in the $H_2$ matrix. For a desired code rate $R_p$ to be obtained from puncturing the mother code of rate $R_L$, the number of puncturing symbols is $p = N\left(1 - R_L/R_p\right)$, where $N$ is the code length and $R_L \le R_p \le R_H$. Any desired code rate can be achieved by first puncturing nodes from the *1*-SR matrix, then from the *2*-SR matrix and so on. Thus the codes of different rates can be applied to IR Hybrid-ARQ systems.

## IV. Efficient Encoder Implementation

In this section we show that E[2]RC codes can be encoded in linear time. We start by presenting an efficient shift register based technique that can be applied to other similar block codes as well. First, we will explain the case when $N_v(2) = M-1$. For the parity-check matrix $H = [H_1 \mid H_2]$ of an E[2]RC code obtained from the proposed construction algorithm, let a codeword $c = [m \mid p]$, where $m$ is the systematic symbols, and $p$ is nonsystematic symbols. Then, we have $H \cdot c^T = [H_1 \mid H_2] \cdot [m \mid p]^T = H_1 m^T + H_2 p^T = 0$. Let $s^T = H_1 m^T$, then we have $H_2 p^T = H_1 m^T = s^T$. Since $H_1$ is sparse $s$ can be found efficiently.

$$H_2 \cdot p^T = \begin{bmatrix} h_{11} & h_{12} & \cdots & h_{1M} \\ h_{21} & h_{22} & \cdots & h_{2M} \\ h_{31} & h_{32} & \cdots & h_{3M} \\ \vdots & \vdots & \ddots & \vdots \\ h_{M1} & h_{M2} & \cdots & h_{MM} \end{bmatrix} \begin{bmatrix} p_1 \\ p_2 \\ p_3 \\ \vdots \\ p_M \end{bmatrix} = \begin{bmatrix} s_1 \\ s_2 \\ s_3 \\ \vdots \\ s_M \end{bmatrix}.$$

Let $H_2 = \left(h_{i,j}\right)_{1\le i,j\le M}$, then $s_i = \sum_{j=1}^{M} h_{ij} p_j = \sum_{j=1}^{i-1} h_{ij} p_j + p_i$ since $h_{ij} = 1$ for $i = j$ and $h_{ij} = 0$ for $i < j$ (since $H_2$ is lower triangular) in the construction of the $E^2RC$ codes. Since all nodes in $H_2$ are degree-2, the elements between the two entries of the sequence are 0. This means that for $1 \le j \le \gamma(1)$,

$$h_{ij} = \begin{cases} 1, & i = j \quad or \quad i = j + \gamma(1) \\ 0, & otherwise \end{cases}.$$

Then we have

$$p_i = \begin{cases} s_i, & for\ 1 \le i \le \gamma(1) \\ s_i + \sum_{j=1}^{i-1} h_{ij} p_j, & for\ \gamma(1) + 1 \le i \le M \end{cases}.$$

The above results tell us that we can get $p_i$ for $1 \le i \le \gamma(1)$ directly from $s_i$. By using the obtained $\gamma(1)$ $p_i$'s, we can get $p_i$ one by one for $\gamma(1) + 1 \le i \le M$, which enables us to implement the $E^2RC$ encoder by using $\gamma(1)$ shift registers. The following example illustrates the encoding method.

*Example 3*: For $M$=7, we can construct $H_2$ matrix as follows:

$$H_2 \cdot p^T = \begin{bmatrix} 1 & 0 & 0 & 0 & 0 & 0 & 0 \\ 0 & 1 & 0 & 0 & 0 & 0 & 0 \\ 0 & 0 & 1 & 0 & 0 & 0 & 0 \\ \boxed{1 \quad 0 \quad 0} & & & 1 & 0 & 0 & 0 \\ 0 & \boxed{1 \quad 0 \quad 0} & & & 1 & 0 & 0 \\ 0 & 0 & \boxed{1 \quad 1 \quad 0} & & & 1 & 0 \\ 0 & 0 & 0 & \boxed{0 \quad 1 \quad 1} & & & 1 \end{bmatrix} \begin{bmatrix} p_1 \\ p_2 \\ p_3 \\ p_4 \\ p_5 \\ p_6 \\ p_7 \end{bmatrix} = \begin{bmatrix} s_1 \\ s_2 \\ s_3 \\ s_4 \\ s_5 \\ s_6 \\ s_7 \end{bmatrix}.$$

Simplifying we get: $p_1 = s_1$, $p_2 = s_2$, $p_3 = s_3$, $p_1 + p_4 = s_4$, $p_2 + p_5 = s_5$, $p_3 + p_4 + p_6 = s_6$, and $p_5 + p_6 + p_7 = s_7$. Then, we can obtain $p_i$'s by using $p_j$'s, where $j < i$: $p_1 = s_1$, $p_2 = s_2$, $p_3 = s_3$, $p_4 = p_1 + s_4$, $p_5 = p_2 + s_5$, $p_6 = p_3 + p_4 + s_6$, and $p_7 = p_5 + p_6 + s_7$. We only need $\gamma(1) = 3$ memory elements for the encoder in Fig. 3. The coefficients for multiplication in Fig. 3 can be obtained from the sliding windows highlighted as squares in the matrix equation. For this reason, we will refer to this encoding method as sliding window method. The coefficient $g_i$'s are time varying. Assuming that the window starts from the first row at initial time $t$=0, $g_0$ will be on during $t$=3~5, $g_1$ will be on during $t$=5~6, and $g_2$ will be on at $t$=6. ■

From the Example 3, we can generalize the shift-register encoder implementation of E[2]RC codes. The encoder can be represented as division circuit as shown in Fig. 3. The division circuit can be specified by a generator polynomial $g(x) = g_0 + g_1 x + g_2 x^2 + \cdots + g_{\gamma(1)-1} x^{\gamma(1)-1} + x^{\gamma(1)}$. By observing the matrix $H_2$, we can obtain the coefficients of the polynomial. As in Fig. 4, consider the window of size $w$. As we slide the window from the first row to the last row, we can get parity-check equations one by one. The coefficients in the window will change or stay between 0 and 1 for each row. If we trace the time-varying coefficients, then we can implement the shift-register encoder of Fig. 3.

We set the window size $w$ as $\gamma(1)$ since the largest distance between nonzero elements in a row of $H_2$ is $\gamma(1)$. The window size can be set differently for other codes. In the sliding window, the first entry corresponds to $g_0$, and the last entry to $g_{\gamma(1)-1}$. Let us define the time, $t = 0$ when the window starts from the first row. The initial status of coefficients is 0. In the code construction, note that $g_i$ can exist only if $i = \gamma(1) - \gamma(k)$ for $1 \le k \le d$. In other words, we only have to consider $d$ coefficients and other than those are all zero. For a such coefficient $g_i$, it is on at time $t = S_k$, and will last until the window reach the last row ($t = S_d$) if there is a connection for $k$-SR matrix in the last row. Otherwise, it will be off at the last row. Fig. 5 shows the timing diagram of coefficients. From Observation 1, note that there is a connection for $k$-SR matrix in the last row if $\gamma(k)+S_k-S_d=1$ and no connection if the value is 0. Then, the coefficients of the generator polynomial $g(x)$ can be represented as

$$g_i=\sum_{k=1}^{d}\delta(i-\gamma(1)+\gamma(k))\{u(t-S_k)-\delta(\gamma(k)+S_k-S_d)\cdot u(t-S_d)\},$$

where we define the unit step function as follows:

$$u(t)=\begin{cases}1, & t\geq 0\\ 0, & t<0.\end{cases}$$

For the above Example 3 when $M$ =7, $g_0=u(t-3)-u(t-6)$, $g_1=u(t-5)$, $g_1=u(t-6)$. As mentioned earlier, the proposed sliding window encoding method can be applied to other block codes if the nonsystematic part of their parity-check matrix has lower-triangular form as shown in Fig. 4. In fact, the window size can be lowered if the lower-triangular form in Fig. 4 has lower-triangular 0's in it, which can be attempted by column and row permutation for a given parity-check matrix.

Another way to implement the encoder of the proposed $E^2$RC codes is by using a simple iterative erasure decoder. Recall that all the nodes in *k*-SR matrix can be recovered in *k* iterations of erasure decoding since they are all *k*-SR nodes. For the proposed codes, even if all the parity bits are erased, we can obtain the exact parity bits within ($d$+1) iterations using a simple erasure decoder or a general message-passing LDPC decoder as long as the systematic bits are known exactly (this is the case at the encoder). In a transceiver system, this can be a big advantage in terms of complexity. We only need to provide an LDPC decoder for both encoding and decoding, and do not need any extra encoder.

Even though we may not be able to use the shift-register implementation of sliding-window method for the encoder when $N_v(2)<M-1$, we can easily apply the efficient encoding method proposed in [1]. Following the notation in [1], let the parity-check matrix $H$ be represented as $H=\begin{bmatrix}A & B & C\\ D & E & F\end{bmatrix}$. Then, $\begin{bmatrix}A\\ D\end{bmatrix}$ is the systematic part of the $E^2$RC codes , $\begin{bmatrix}B\\ E\end{bmatrix}=L$ is the submatrix of the nonsystematic part consisting of nodes with degree higher than two and $\begin{bmatrix}C\\ F\end{bmatrix}=T$ is the submatrix of the nonsystematic part consisting of degree two nodes. For E$^2$RC codes, we know the exact sequence of the matrix *T*. Furthermore, the matrix *C* is a lower triangular with ones on the diagonal. Thus preprocessing is not required for putting the matrix in the form used in [1]. This makes it easy to apply the efficient encoding techniques in [1] to E$^2$RC codes.

## V. SIMULATIONS

In this section we present simulation results of E$^2$RC codes and the compare their performance with eIRA codes and general irregular LDPC codes. We consider rate-1/2 mother codes with block length of 1200. For a fair comparison we use the degree distributions presented in [5] for rate-1/2 codes:

$$\lambda(x) = 0.30780x + 0.27287x^2 + 0.41933x^6$$
$$\rho(x) = 0.4x^5 + 0.6x^6.$$

However, the actual degree distributions for the E$^2$RC codes are slightly different to compensate for the right degree of $H_2$. These are given below.

$$\lambda(x) = 0.00025 + 0.30199x + 0.27073x^2 + 0.42702x^6$$
$$\rho(x) = 0.40685x^5 + 0.55054x^6 + 0.01815x^7 + 0.01361x^8 + 0.00504x^9 + 0.00278x^{10} + 0.00303x^{11}.$$

The progressive edge growth (PEG) algorithm in [12] is applied to design the systematic matrix $H_1$ to improve girth characteristics for both eIRA codes and E$^2$RC codes. The PEG algorithm was also used to generate the general irregular LDPC codes. First, we consider random puncturing for the puncturing strategy for eIRA codes and general irregular LDPC codes. The puncturing performance comparisons between eIRA codes and E$^2$RC codes are shown in Fig. 6. From Fig. 6, the E$^2$RC codes show more powerful puncturing performance at higher code rates. At a rate of 0.8 and BER = $10^{-5}$, E$^2$RC codes outperform eIRA codes by over 0.8dB. A similar performance gap was observed in the comparisons with general irregular LDPC codes (the curve is omitted due to lack of space).

Next, we apply the puncturing algorithm proposed in [8-9] to eIRA codes and general irregular LDPC codes. As mentioned earlier, this puncturing algorithm has a limit on the number of low-SR nodes that it can find. In fact, the puncturing algorithm in [8-9] assigns 300 nodes as *1*-SR nodes, and cannot find further *k*-SR nodes ($k \geq 2$) if we try to maximize the number of *1*-SR nodes. To get a high rate ($R = 0.7, 0.8, 0.9$) in eIRA codes, we puncture randomly after the puncturing limitation of rate 0.67, which destroys the previous tree structure of *1*-SR nodes resulting in poor performance. To increase the number of variable nodes that can be punctured for eIRA codes, one can impose a limitation on the number of the lower-SR nodes when the puncturing algorithm in [8-9] is applied, thus trading off fewer 1-SR nodes for more number of *2*-SR and *3*-SR nodes. In this case, however, the puncturing performance for lower rate is worse than the case when *1*-SR nodes are maximized. For general irregular LDPC codes, we can find 389 *1*-SR nodes, 45 *2*-SR nodes, 2 *3*-SR nodes, so the maximum puncturing rate is 0.785. Above the puncturing limit, we apply random puncturing to get higher rates. The puncturing performance of eIRA codes and general irregular LDPC codes with the puncturing algorithm in [8-9] are shown in Fig. 7 and 8. Even with the best effort intentional puncturing algorithm in [8-9], the $E^2$RC codes show better puncturing performance across the entire range of rates, especially at higher rates. For code rate of 0.9 and BER of $10^{-5}$ the $E^2$RC codes outperform the eIRA codes and general irregular LDPC codes by 0.7dB and 1.5 dB respectively.

For practical purposes, designing a low rate $E^2$RC code and providing a wide range of rates by puncturing are useful. There are other methods to lower the rates such as extending and shortening. However, these methods often increase hardware complexity or the performance of lower rate code may not be good enough. On the other hand, punctured low rate ($R < 0.5$) standard irregular mother codes have bad performance at high puncturing rates. The $E^2$RC codes show relatively less performance degradation when punctured as compared to other LDPC codes. For $E^2$RC codes, all the degree-2 nodes in the parities can be punctured. As an example, we consider a rate-0.4 mother code of which degree distributions are optimized in AWGN channel (edge perspective):

$$\lambda(x) = 0.29472x + 0.25667x^2 + 0.44861x^9$$
$$\rho(x) = x^5.$$

Since we assign all the degree-2 nodes to parities and higher degree nodes to messages, 88.4% of the parities are degree-2 nodes and the remaining 11.6% of the parities are degree-3 nodes from the above degree distributions. Thus, the structure of E$^2$RC codes is changed from the original one, and the E$^2$RC codes can achieve rate of 0.85 since all the degree-2 nodes can be punctured. For rate-0.4 mother code with $N = 2000$, $K = 800$, and $N_v(2) = 1061$, we have the depth $d$ = 4, and $\gamma(1) = 600$, $\gamma(2) = 300$, $\gamma(3) = 150$, $\gamma(4) = 11$. In addition, the E$^2$RC codes can have perfect right degree concentration at degree 6. We apply the PEG algorithm to generate matrix other than degree-2 parities. To compare the puncturing performance, the general irregular LDPC codes with the same degree distributions as above are generated by using the PEG algorithm. The best-effort puncturing algorithm in [8-9] is applied to the general irregular LDPC codes. The maximum achievable rate of this general irregular LDPC code is 0.69 with puncturing algorithm in [8-9]. So, after the limit we apply random puncturing. The puncturing performance comparison between E$^2$RC codes and general irregular LDPC codes is depicted in Fig. 9 and 10. In Fig. 9 and 10, the E$^2$RC codes show good performance over a wide range of rates 0.4~0.85. At a BER of $10^{-5}$ in Fig. 9, the E$^2$RC codes outperform the general irregular LDPC codes by 1.0dB and 2.7dB at rate 0.8 and 0.85, respectively. The same trend can be observed in FER performance in Fig. 10.

## VI. CONCLUSIONS

We have proposed a new class of codes called E$^2$RC codes that have several desirable features. First, the codes are efficiently encodable. We have presented a shift-register based implementation of the encoder which has low-complexity and demonstrated that a simple erasure decoder can be used for the encoding of these codes. Thus, we can share a message-passing decoder for both encoding and decoding if it is applied to transceiver systems which require an encoder/decoder pair. Second, we have shown that the nonsystematic part of the parity-check matrix are cycle-free, which ensures good code characteristics. From simulations, the performance of the E$^2$RC codes (mother codes) is as good as that of eIRA codes and other irregular LDPC codes. Third, the E$^2$RC codes show better performance under puncturing than other irregular LDPC codes and eIRA codes in all ranges of code rates and are particularly good at high rates. Finally, the E$^2$RC codes can provide good performance over a wide range of rates even when they are designed for rates lower than 0.5. We believe that these characteristics of E[2]RC codes are valuable when they are applied to IR Hybrid-ARQ systems.

## APPENDIX

*Proof of Fact 1*: From the definition, $M$ should be $2^{d-1} < M \le 2^d$. By definition, $M$ can be represented by $M = 2\cdot\gamma(1) + R_1$, where $R_1$ is the remainder when M is divided by 2, i.e., $R_1 = 0$ or 1. Then, we have

$$M - \gamma(1) = \gamma(1) + R_1 = 2\cdot\gamma(2) + R_2$$

$$M - \gamma(1) - \gamma(2) = \gamma(2) + R_2 = 2\cdot\gamma(3) + R_3$$

$$\cdots$$

$$M - \gamma(1) - \gamma(2) - \cdots - \gamma(d-1) = \gamma(d-1) + R_{d-1} = 2\cdot\gamma(d) + R_d \qquad \text{(a)}$$

In the above equations, the remainders can be $R_1, R_2, ..., R_d = 0$ or 1. From the equations above, we also have

$$M + R_1 = 2\cdot(\gamma(1) + R_1) = 2\cdot(2\cdot\gamma(2) + R_2)$$

$$M + R_1 + 2\cdot R_2 = 2^2\cdot(\gamma(2) + R_2) = 2^2\cdot(2\cdot\gamma(3) + R_3)$$

$$\cdots$$

$$M + R_1 + 2\cdot R_2 + \cdots 2^{d-2}\cdot R_{d-1} = 2^{d-1}\cdot(\gamma(d-1) + R_{d-1}) = 2^d\cdot\gamma(d) + 2^{d-1}\cdot R_d \qquad \text{(b)}$$

$$M + R_1 + 2\cdot R_2 + \cdots 2^{d-1}\cdot R_d = 2^d\cdot(\gamma(d) + R_d) \qquad \text{(c)}$$

In equation (b), the LHS is strictly greater than $2^{d-1}$ from the range of *M*. So, $\gamma(d) \ge 1$ in RHS since $R_d = 0$ or 1. On the other hand, $\gamma(d) + R_d$ in equation (c) has to be 1 since the sum of the LHS of (c) is at most $2^{d+1}$ - 1. Thus, we conclude that $\gamma(d) = 1$ and $R_d = 0$. Then, from (a), we have $\gamma(1) + \gamma(2) + \cdots + \gamma(d) = M - 1$.

Now, note that $\gamma(1) \ge \gamma(2) \ge \cdots \ge \gamma(d)$ and since $\gamma(d) = 1$, therefore $\gamma(k) \ge 1$ for $1 \le k \le d$.

■

*Proof of Lemma 1:* Note that $H_2$ is lower-triangular with ones on the diagonal. Therefore in the case when $k = 1$, since $h_{1,j_1} = D^{j_1}\left(1 + D^{\gamma(1)}\right)$ for $0 \le j_1 \le \gamma(1) - 1$ we have a set of columns whose first entry is on the diagonal. Therefore the first entry of these columns is connected to a row of degree-1 and the lemma holds for $k=1$. Now consider $k \geq 2$. The $j_k$-th column in the $k$-SR matrix has a sequence is given by

$$\begin{aligned} h_{k,j_k} &= D^{j_k+S_{k-1}}\left(1+D^{\gamma(k)}\right) \\ &= D^{j_k+S_{k-1}} + D^{j_k+S_k}, \text{ where } 0 \leq j_k \leq \gamma(k)-1. \end{aligned}$$

We shall demonstrate that the first entry of $h_{k,j_k}$ is connected to a column in the $l$-SR matrix for $1 \leq l < k$. An immediate consequence of the lower-triangular nature of $H_2$ is that $h_{k,j_k}$ can only be connected to the second entry of $h_{l,j_l}$, the $j_l$-th column in the $l$-SR matrix. Suppose that the second entry of the $j_l$-th column in the $l$-SR matrix is connected to the first entry of the $j_k$-th column in the $k$-SR matrix. This implies $j_l + S_l = j_k + S_{k-1}$. Clearly $j_l = j_k + S_{k-1} - S_l \geq 0$ since $k > l$ and $0 \leq j_k \leq \gamma(k)-1$. We shall now show that $j_l = j_k + S_{k-1} - S_l \leq \gamma(l)-1$. This means that for a given $j_k$, it is possible to find a unique column $j_l$ belonging to the $l$-SR matrix to which it is connected. From the proof of Fact 1, we have $S_i = M - \gamma(i) - R_i$, where $R_i = 0$ or 1. Since $S_{k-1} + \gamma(k) = S_k$ and $S_k \leq S_d$, we have

$$\begin{aligned} j_l = j_k + S_{k-1} - S_l &\leq \gamma(k) - 1 + S_{k-1} - S_l \\ &= S_k - S_l - 1 \\ &\leq S_d - S_l - 1 \\ &= M - \gamma(d) - R_d - \left(M - \gamma(l) - R_l\right) - 1 \\ &= \gamma(l) - 1 - R_l - 1 \\ &\leq \gamma(l) - 1. \end{aligned}$$

Therefore, for a given $j_k$, we can find a corresponding $j_l$ in the $l$-SR matrix for $1 \leq l < k$. Note that the first entry of $j_k$ is connected to the corresponding $j_l$. Since the matrix is lower-triangular, this entry cannot have any connection with a $m$-SR matrix where $m > k$. Therefore this particular row has degree exactly $k$. This concludes the proof. ■

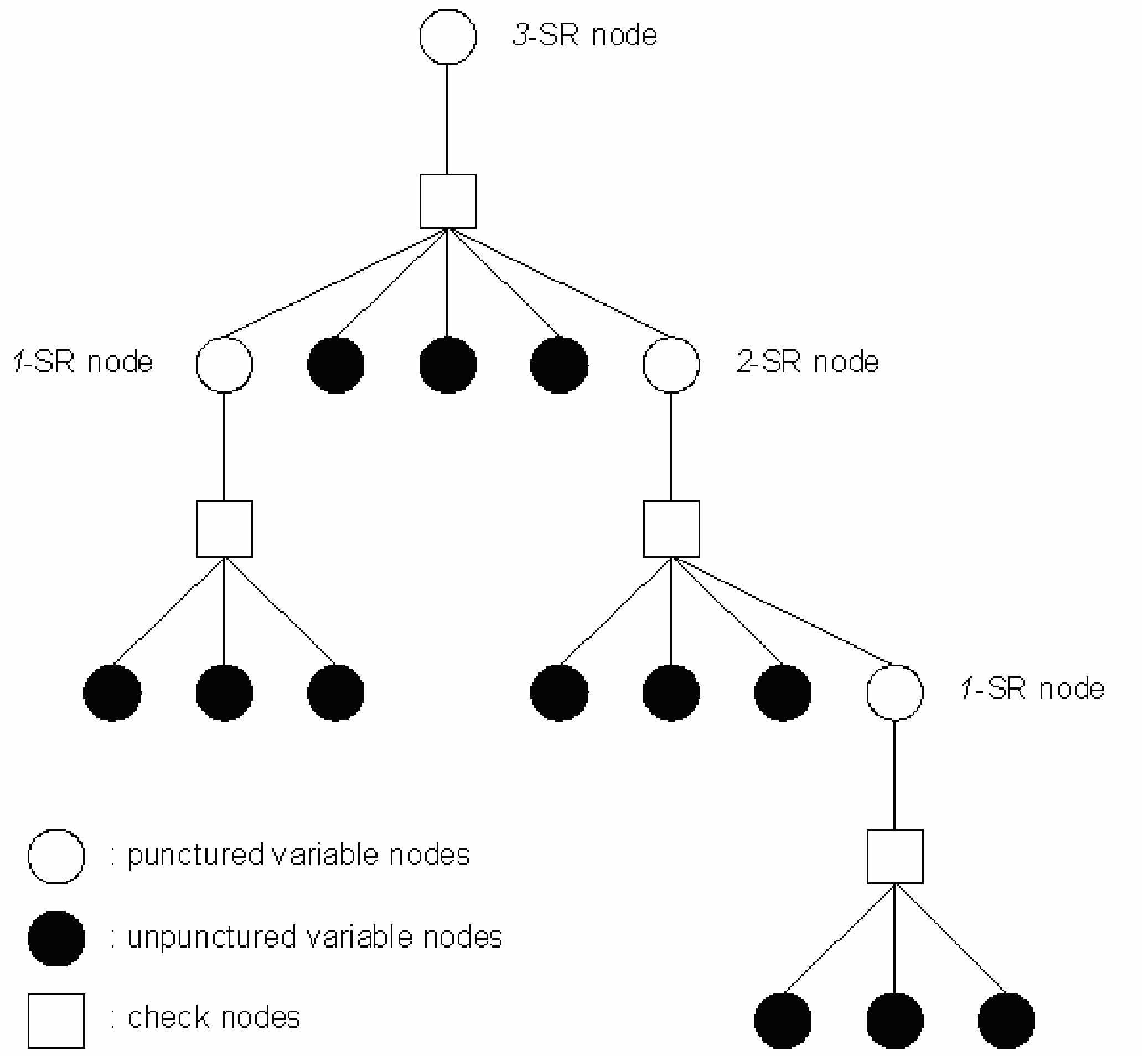

Fig. 1. *k*-SR node in a graph

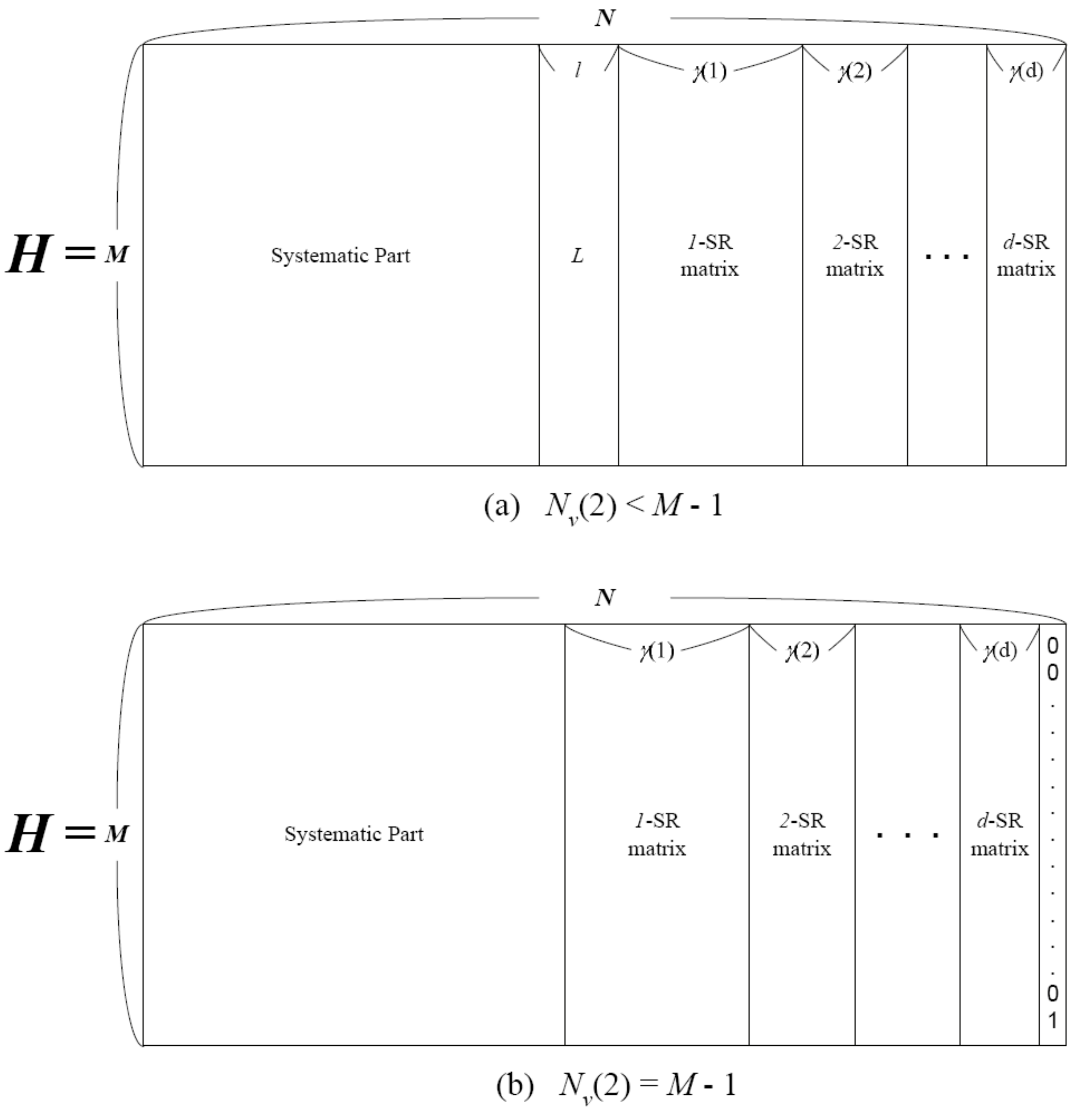

Fig. 2. Construction of the Parity-check matrix of the proposed codes.

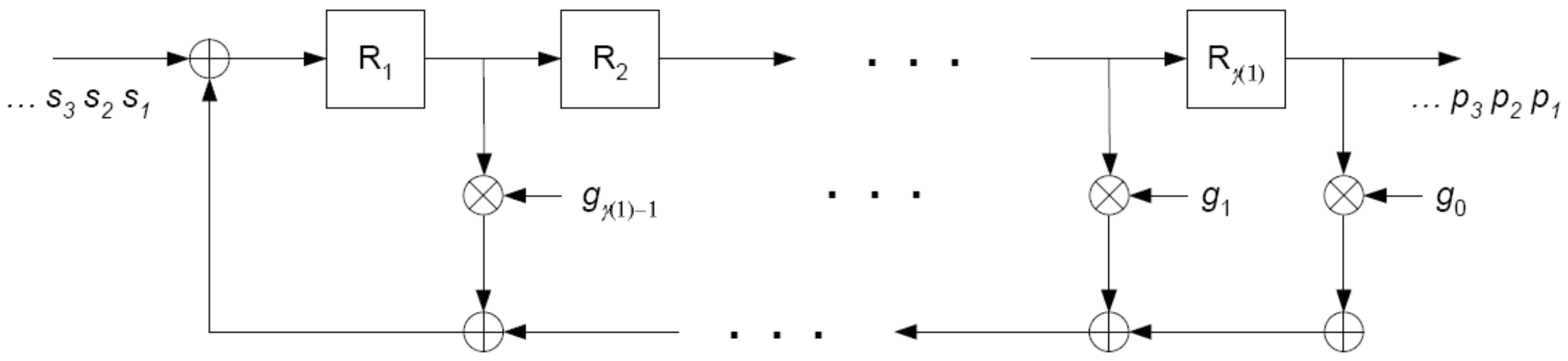

Fig. 3. An example of shift-register implementation of $E^2RC$ codes.

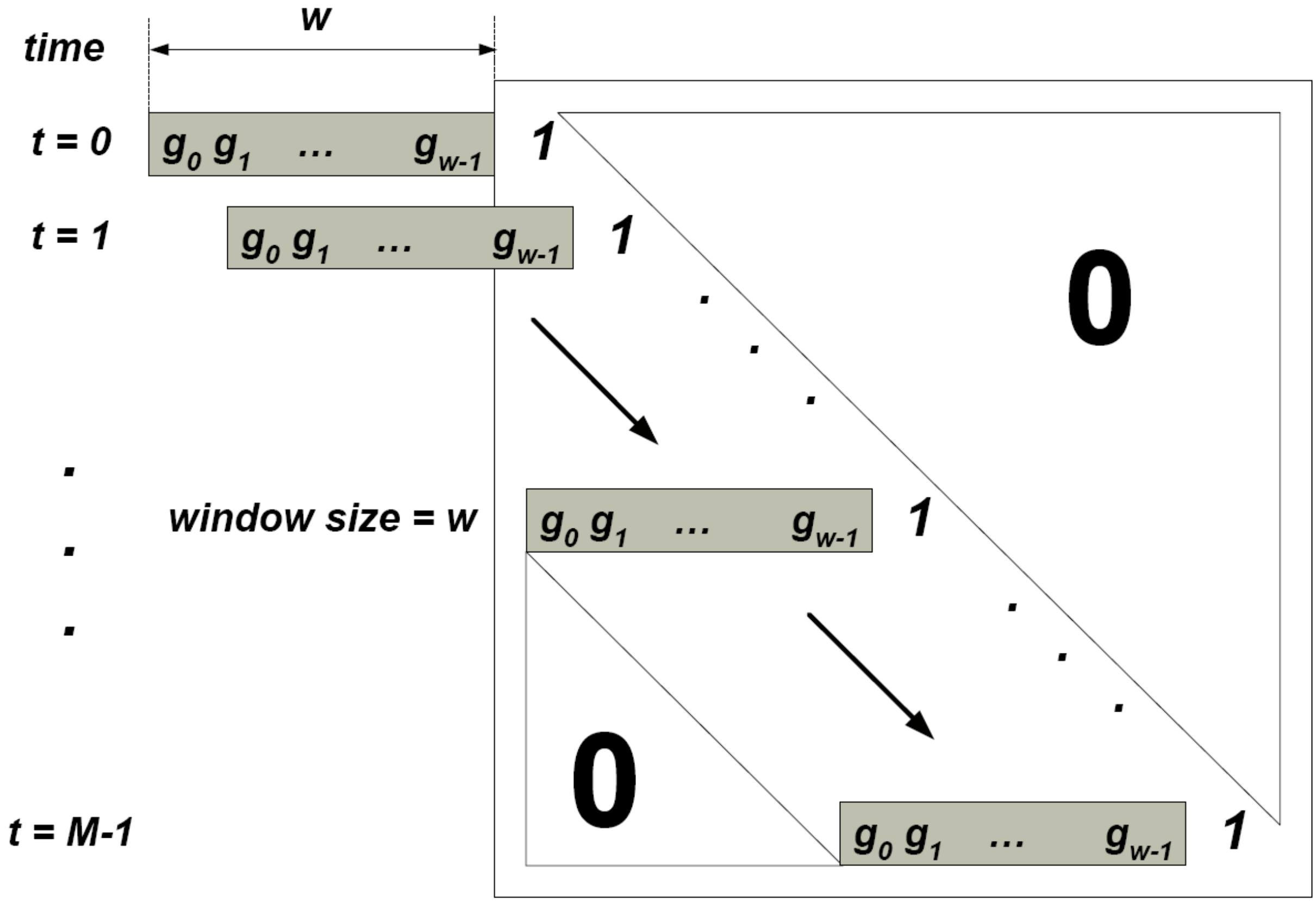

Fig. 4. Nonsystematic part of a parity-check matrix for applying sliding window encoding method.

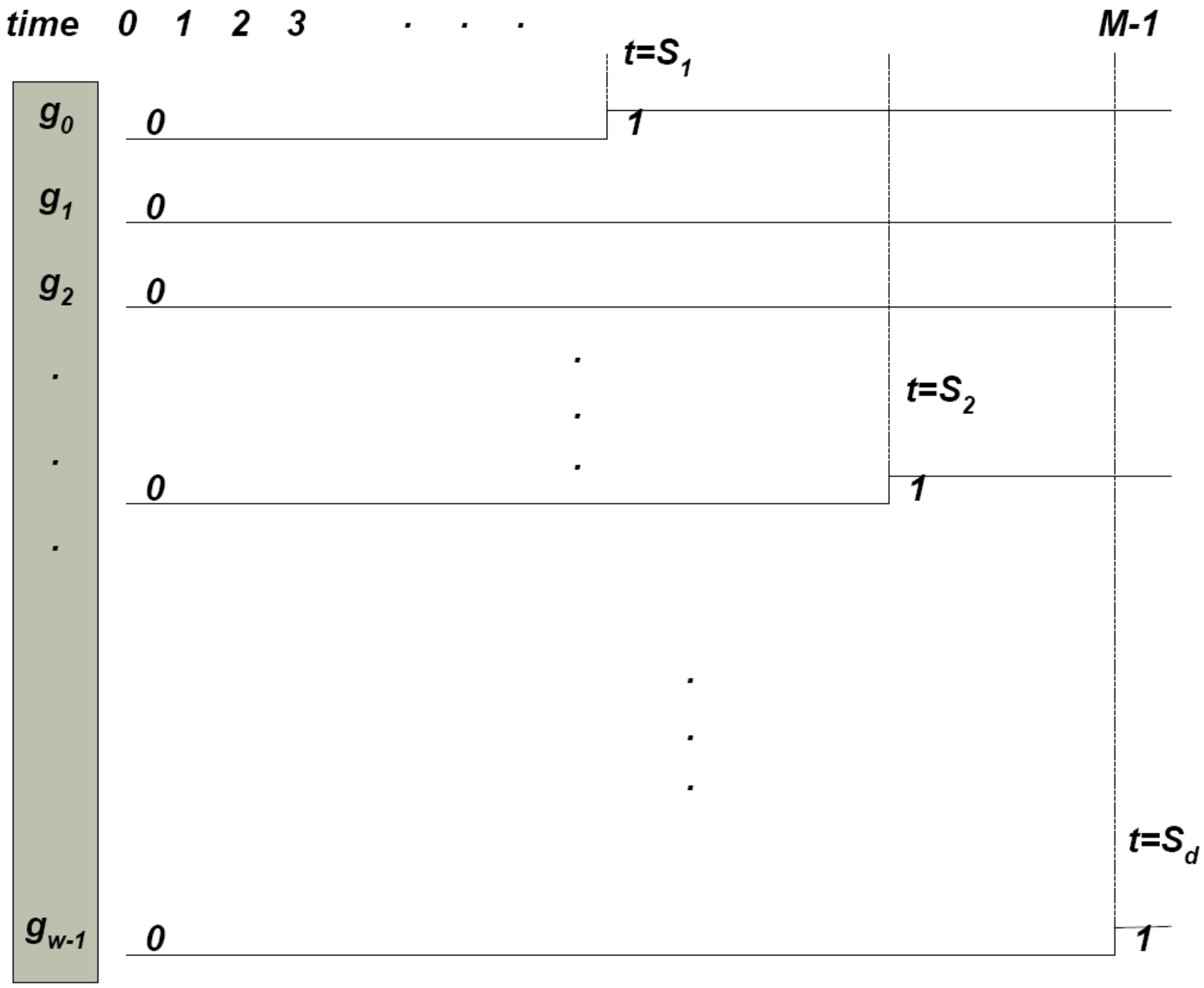

Fig. 5. Timing diagram of coefficients of sliding window encoder.

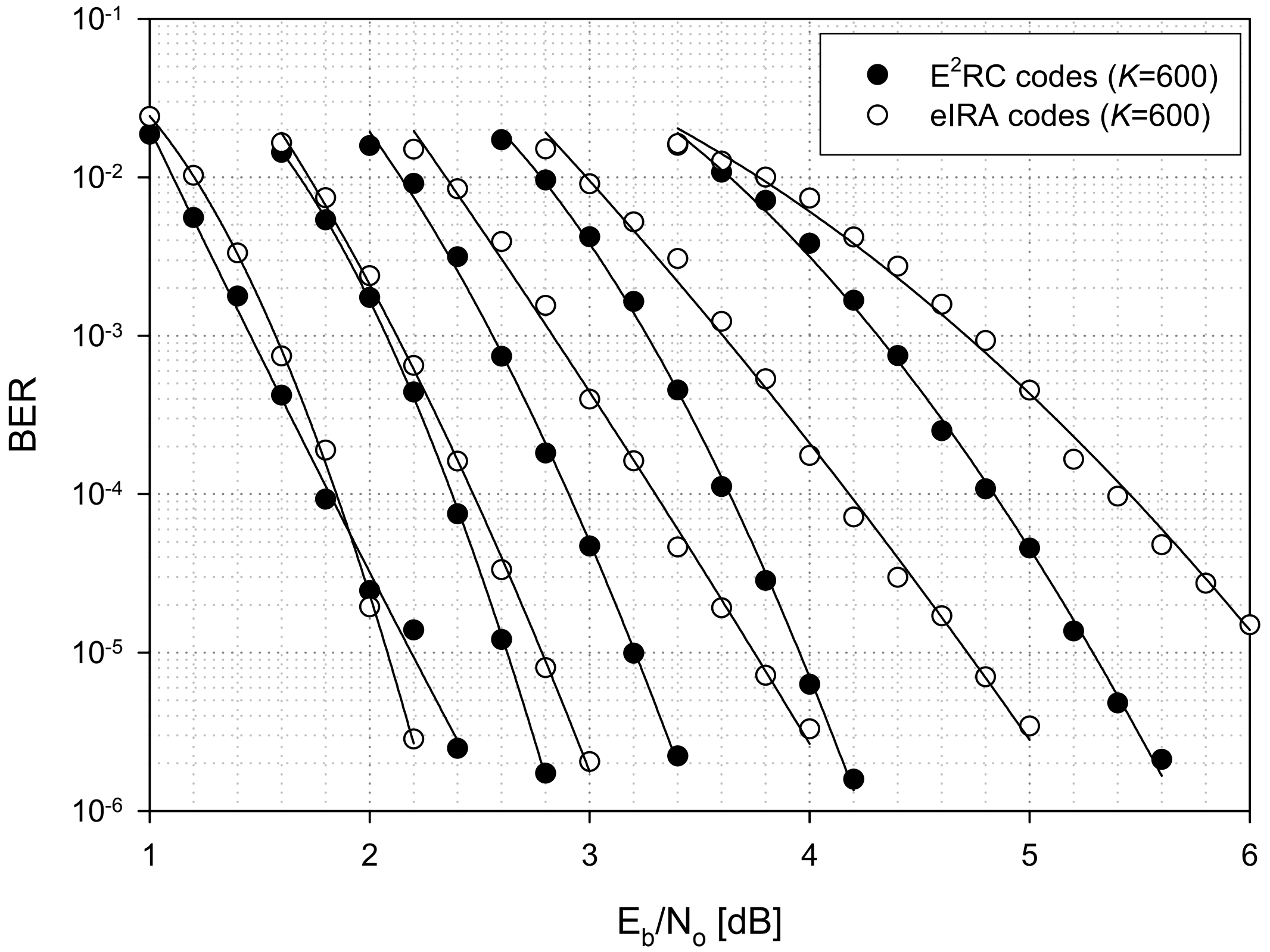

Fig. 6. The puncturing BER performance comparison between E$^2$RC codes and eIRA codes with random puncturing. Filled circles are for E$^2$RC codes and unfilled circles are for eIRA codes. Rates are 0.5 (mother codes), 0.6, 0.7, 0.8, and 0.9 from left to right.

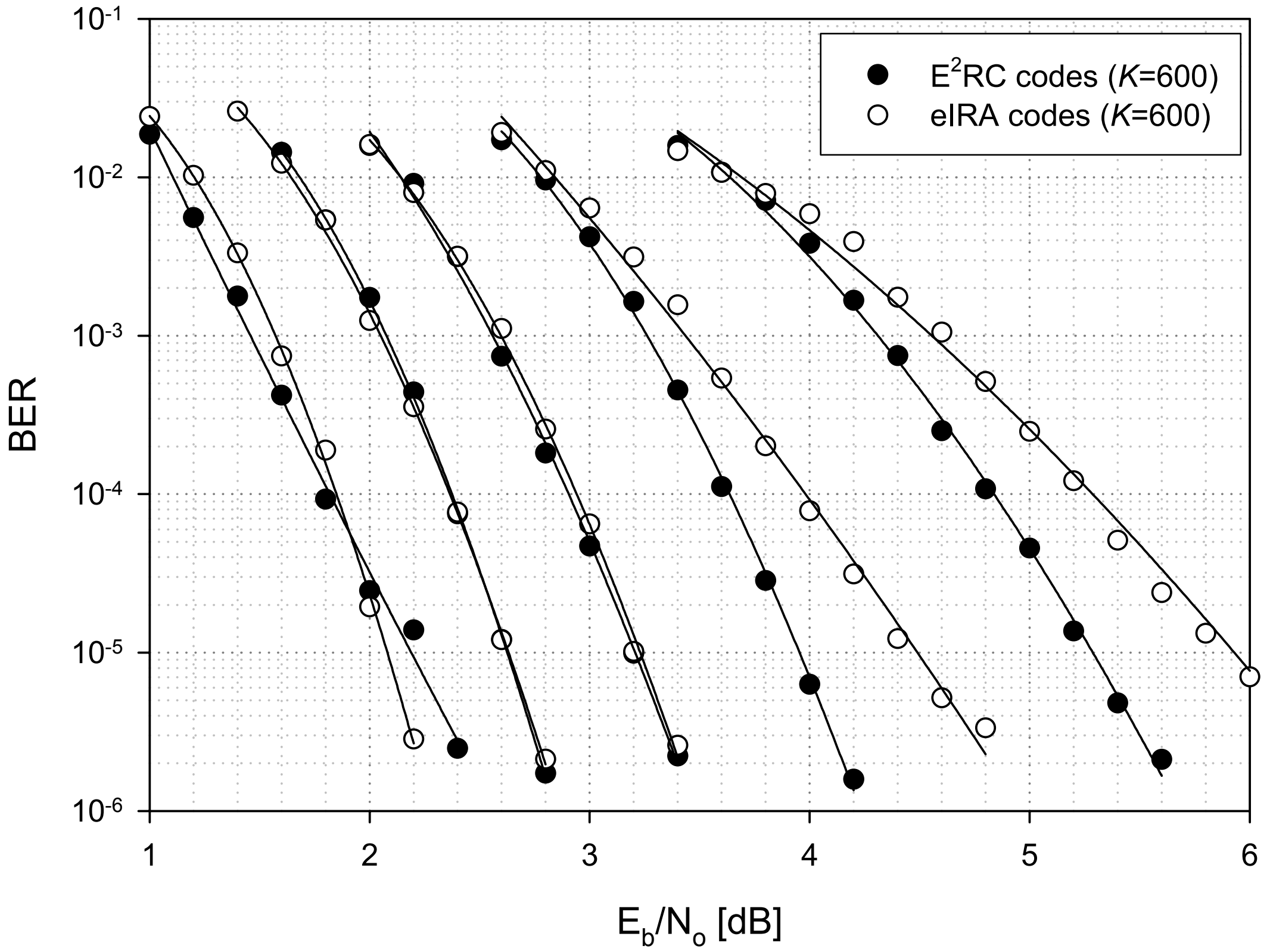

Fig. 7. The puncturing BER performance comparison between $E^2$RC codes and eIRA codes with puncturing algorithm in [8-9]. Filled circles are for $E^2$RC codes and unfilled circles are for eIRA codes. Rates are 0.5 (mother codes), 0.6, 0.7, 0.8, and 0.9 from left to right

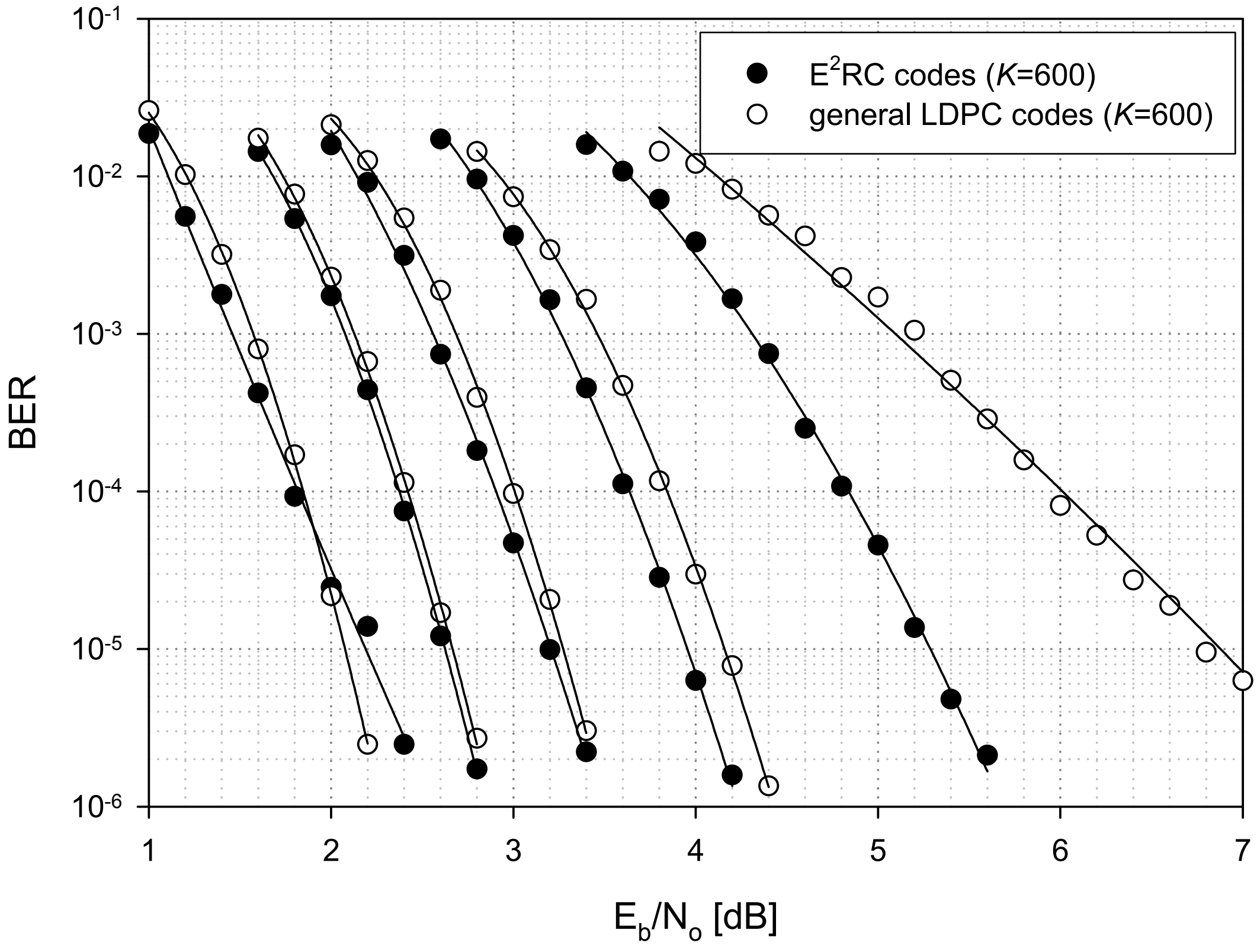

Fig. 8. The puncturing BER performance comparison between E$^2$RC codes and general irregular LDPC codes with puncturing algorithm in [8-9]. Filled circles are for E$^2$RC codes and unfilled circles are for general irregular LDPC codes. Rates are 0.5 (mother codes), 0.6, 0.7, 0.8, and 0.9 from left to right.

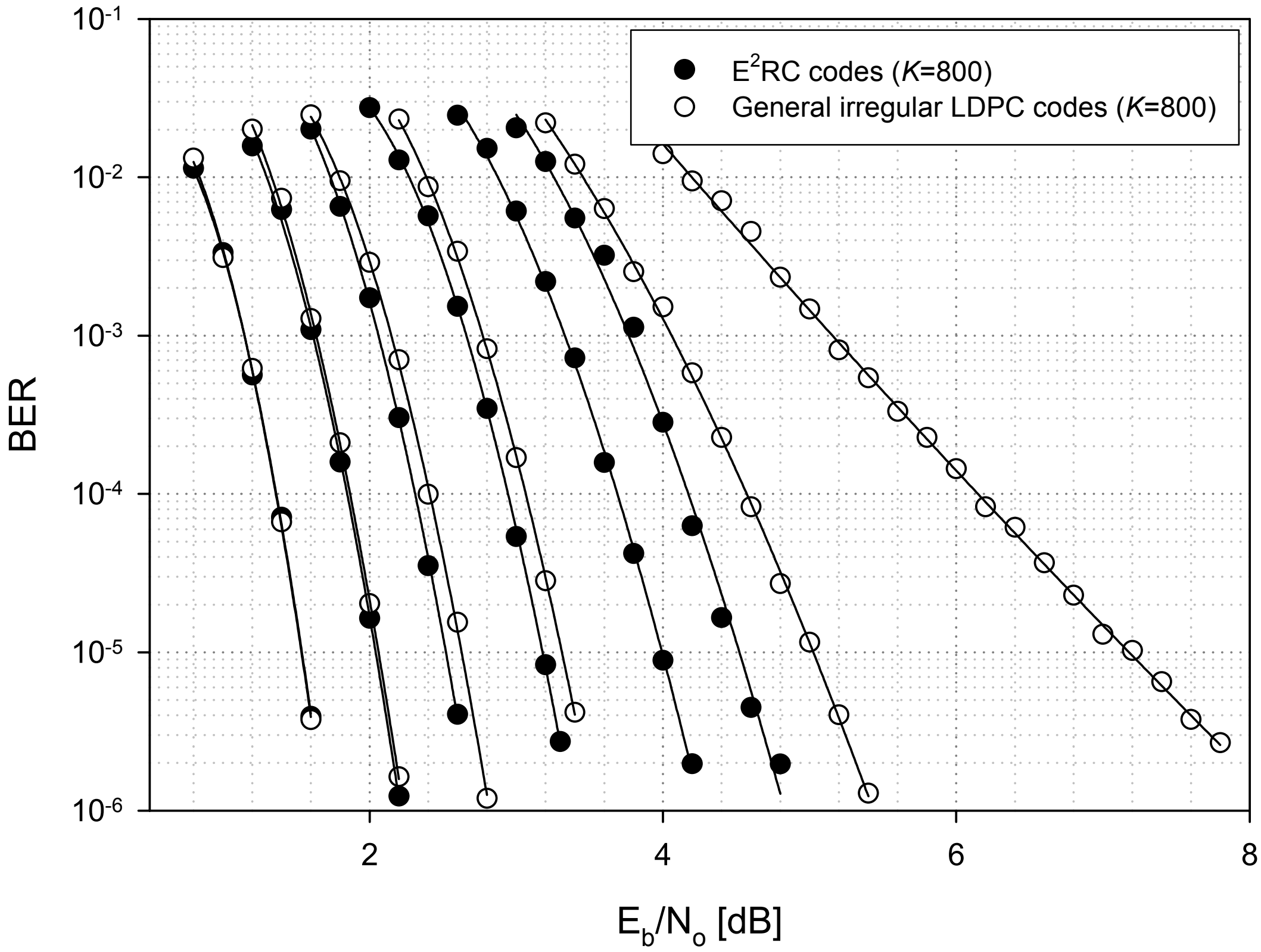

Fig. 9. The puncturing BER performance comparison between $E^2$RC codes and general irregular LDPC codes with puncturing algorithm in [8-9]. Filled circles are for $E^2$RC codes and unfilled circles are for general irregular LDPC codes. Rates are 0.4 (mother codes), 0.5, 0.6, 0.7, 0.8, and 0.85 from left to right.

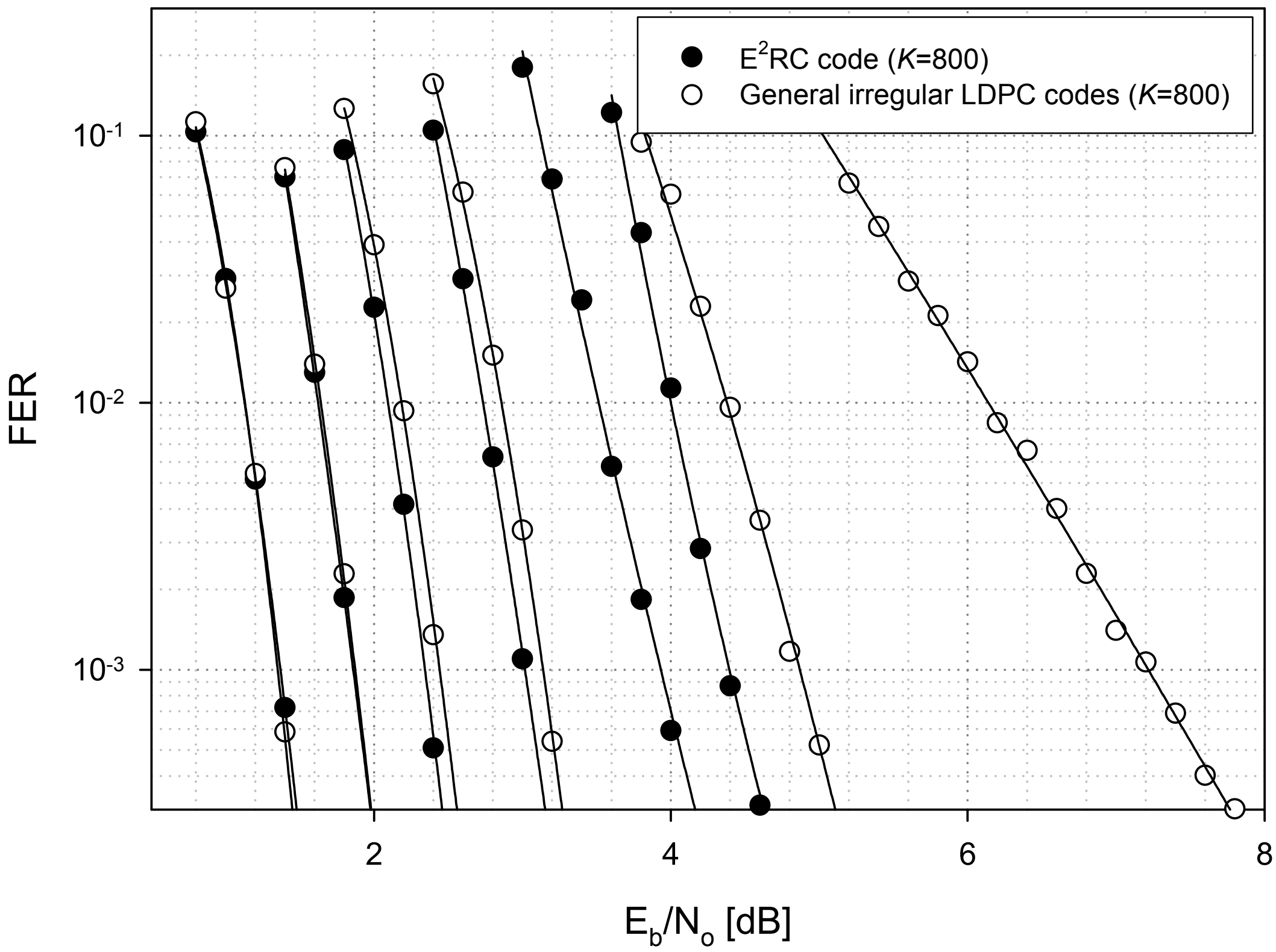

Fig. 10. The puncturing FER performance comparison between E[2]RC codes and general irregular LDPC codes with puncturing algorithm in [8-9]. Filled circles are for E[2]RC codes and unfilled circles are for general irregular LDPC codes. Rates are 0.4 (mother codes), 0.5, 0.6, 0.7, 0.8, and 0.85 from left to right.